\documentstyle[11pt,aspconf,epsf]{article}

\begin{document}

\title{Dynamical Models of the Milky Way\footnotemark}
\footnotetext{
	Invited talk presented at the meeting ``Formation of the
	Galactic Halo ... Inside and Out'', Tucson, October 9-11,
	1995, held in honor of the 65th birthday of George Preston.
	To appear in the ASP Conference Series, eds.\ Sarajedini A.\
	\& Morrison H.}

\author{Walter Dehnen and James Binney}
\affil{Oxford University, Department of Physics, Theoretical Physics,\\
       1 Keble Road, Oxford OX1 3NP, United Kingdom}

\begin{abstract}
The only way to map the Galaxy's gravitational potential $\Phi({\bf x})$ and the
distribution of matter that produces it is by modelling the dynamics of stars
and gas. Observations of the kinematics of gas provide key information about
gradients of $\Phi$ within the plane, but little information about the structure
of $\Phi$ out of the plane. Traditional Galaxy models {\em assume}, for each of
the Galaxy's components, arbitrary flattenings, which together with the 
components' relative masses yield the model's equipotentials. However, the
Galaxy's isopotential surfaces should be {\em determined\/} directly from the
motions of stars that move far from the plane. Moreover, from the kinematics
of samples of such stars that have well defined selection criteria, one should
be able not only to map $\Phi$ at all positions, but to determine the
distribution function $f_i({\bf x},{\bf v})$ of each stellar population
$i$ studied. These distribution functions will contain a wealth of information 
relevant to the formation and evolution of the Galaxy. An approach to fitting 
a wide class of dynamical models to the very heterogeneous body of available 
data is described and illustrated.  
\end{abstract}

\section{Introduction}
Models of the Milky Way can be usefully divided into three classes: (i)
photometric models, (ii) kinematic models, and (iii) dynamical models.
Models of the first type were pioneered by Kapteyn and Shapley, a more recent
example is given by Bahcall \& Soneira (1980). These models specify the
distribution of stars without saying anything about their motions.
The models of Bienaym\'e et al.\ (1987) and Ratnatunga et al.\ (1989) are 
typical kinematic models: they specify both the distribution and the motions of
stars. However, in these models the velocity distribution at each point must be
somehow specified without fully exploiting Newton's laws of motion. In practice
the Galaxy is decomposed into `components' such as the thin disk, the thick 
disk, the metal-poor halo etc, and at each point in space each component is 
assigned an ellipsoidal, usually Gaussian, velocity distribution. One might 
hope to use the Jeans' equations to couple the ellipsoids assigned at 
neighboring points, but Eddington (1915) already showed that this is feasible
only if the potential is of St\"ackel's special form, which it almost certainly 
is not. So the velocity ellipsoid at each point in each component must be 
arbitrarily assigned at risk being dynamically impossible. In consequence, 
models of this type may have a role to play as representations of observational 
data, but they by no means fully exploit the potential of stellar surveys.

This contribution emphasizes the value of constructing models in which stellar
dynamics plays an integral role, and explains why such models have until
recently been of an extremely limited nature, and why the time is now ripe
for them to come into their own.

\section{Orbits and Integrals}
The basic result of galactic dynamics is that galaxies can be considered to
be made up of orbits in a smooth gravitational potential. From this it follows
that a prerequisite for dynamical modelling is an understanding of such orbits.
In the simplest circumstances orbits in a steady three-dimensional potential
are characterized by three `isolating integrals,' that is functions 
$I_i({\bf x},{\bf v})$ of position and velocity that are constant along any
orbit. Through the work of Lindblad (1933), Contopoulos (1963), and H\'enon \& 
Heiles (1964) it has long been clear that most orbits in typical axisymmetric 
galactic potentials admit three such integrals. Two integrals are available
from elementary mechanics, $I_1\,{=}\,E$, the star's energy, and $I_2\,{=}\,
L_z$, the star's angular momentum about the potential's symmetry axis. The 
problem is that, for most systems of interest, no general formula exists for a
third integral, $I_3$. By Jeans' theorem, the simplest assumption one can make
on an equilibrium system is that the probability density $f({\bf x},{\bf v})$,
that a given star will be found at the phase-space point $({\bf x},{\bf v})$, 
is a function $f(I_1,I_2, I_3)$ of three isolating integrals. Hence ignorance
of $I_3$ makes it impossible to express a general distribution function (DF)
$f$ in its simplest form.

External galaxies have been modelled dynamically more extensively than has
the Milky Way by the simple stratagem of ignoring $I_3$ (e.g., Binney et al.\
1990). When the observational data are sparse, this procedure can lead to 
acceptable models. Unfortunately, it has been known since Jeans' classic (1915)
paper that this stratagem cannot lead to an acceptable model of the Milky Way,
because it predicts that the velocity dispersions in the radial and vertical
directions must be equal; near the Sun they differ by nearly a factor of 2.

\subsection{The Oort--Lindblad approximation}
Oort (1932) and Lindblad (1933) decomposed the stellar motion into a vertical
oscillation and the motion parallel to the plane. The latter can be further
divided into azimuthal rotation and radial libration. Let $(R,\varphi,z)$ be 
cylindrical polar coordinates, then the difference between a star's energy
$E\,{\equiv}\,\frac{1}{2}v^2\,{+}\,\Phi({\bf x})$ and the energy $E_c\,{+}\,E_R$
associated with motion parallel to the plane provides an approximate form of
$I_3$:
\begin{equation}\label{OL}
	E_z \equiv E-(E_c+E_R)
	= {\textstyle \frac{1}{2}} v_z^2 +\Phi(R,z)-\Phi(R,0).
\end{equation}
Here $E_c(L_z)$ is the energy of a circular orbit with angular momentum $L_z$
and 
\begin{equation}\label{OLtwo}
	E_R\equiv{\textstyle\frac{1}{2}} v_R^2 + \Phi_{\rm eff}(R) - E_c(L_z)
\quad\mbox{where}\quad
	\Phi_{\rm eff}(R)\equiv\Phi(R,0)+\frac{L_z^2}{2R^2}
\end{equation}
is the energy associated with radial librations. This Oort--Lindblad form of
$I_3$ plays a key role in all published determinations of the local mass 
density within the disk (e.g., Bahcall 1984, Bienaym\'e et al.\ 1987, Kuijken
\& Gilmore 1991), and readily explains the Schwarzschild velocity ellipsoid
(e.g., \S4.2 of Binney \& Tremaine 1987). In fact it enables one to explain the
significant skewness of the observed distributions of $v_\varphi$ velocities 
(Cuddeford \& Binney 1994).

However, (\ref{OL}) is {\em only an approximation\/} and its limitations give
rise to the leading uncertainty in the column density of the disk near the
Sun. The problem is that when $E_z$ is used as third integral, one necessarily
finds that $\langle v_Rv_z\rangle\,{=}\,0$ with the result that even away from
the plane the velocity ellipsoid is aligned with the $R$- and $z$-axes.
By contrast, if the Galaxy's potential were spherically symmetric, the velocity
ellipsoid would everywhere align with the direction to the galactic center. We
can be sure that the real situation is intermediate between these two extremes,
leading to a fundamental uncertainty of $\sim$10\%--20\% in the local surface
density (Kuijken \& Gilmore 1989).

This inability of the O--L approximation to yield an unambiguous determination
of the local vertical force $K_z(z)$ and thus the disk's surface density is the
more disappointing as the latter is one of the very few parameters of the 
galactic potential that is primarily determined from the O--L approximation.
Indeed the O--L approximation is applicable only to stars on nearly circular
orbits -- disk stars. These probe the potential only near the plane, and for 
most purposes the interstellar medium (ISM) provides a more readily observed 
probe of the potential near the plane, since its radio-frequency spectral lines
can be observed throughout the disk without hindrance from the dust which
heavily obscures even moderately distant disk stars. Consequently, the Galaxy's
circular-speed curve $v_c(R)$ has long been determined from observations of the
ISM, and these data can in principle be used to determine $K_z$ near the plane
(Merrifield 1993, Malhotra 1994).

The only way to improve substantially on the knowledge of the galactic potential
that we have gleaned from observations of the ISM is to study the dynamics of
stars that move far from the plane and thus are high-velocity stars when in the
solar neighborhood. The O--L approximation does not apply to these stars, and
some other approach to $I_3$ must be developed. 

\subsection[Approaches to $I_3$]{Approaches to \boldmath$I_3$\unboldmath}
How can we obtain adequate models of the orbits of high-velocity stars? One
possibility is directly to integrate the equations of motion. For each initial
condition chosen, this produces a stream of phase-space positions $[{\bf x}
(t_k),{\bf v}(t_k)],k\,{=}\,1,N$. Two tasks must be addressed before a useful
galaxy model can be based on this raw material: (i) characterize the orbits
in some systematic way, such as assigning values of $E,L_z,I_3$; (ii) put
the output data onto some sort of grid so that one can subsequently decide
whether a given orbit will eventually pass through a given point ${\bf x}$,
and, if so, with what velocity ${\bf v}$.

A recent paper by Zhao (1996) exemplifies this approach. Zhao adopts a
simple model of the rotating potential of the Milky Way's bulge and
calculates several hundred orbits in it. He characterizes his orbits by
energy and the time-averages along them of the angular momentum about the
potential's long and spin axes. He uses the orbit's positions and velocities
to determine associated occupation probabilities and velocities in each of
1000 spatial cells.

Merritt \& Fridman (1995) employ a very similar technique to model
elliptical galaxies. The main difference between their work and Zhao's is
that they follow Schwarzschild (1993) in characterizing orbits not by
time-averaged angular momenta, but through their initial conditions. In this
technique all orbits of a given energy are launched from one or more
two-dimensional surfaces. These surfaces are carefully chosen so that (a) any
orbit can be obtained by launching from some point of one of them, and (b)
as far as possible, each point on the surfaces generates a distinct orbit.
$I_2$ and $I_3$ are simply the values taken by two convenient coordinates
for the starting surface(s) at the orbit's launching point.

In Oxford we have developed an entirely different technique which involves
thinking of orbits as three-dimensional surfaces in six-dimensional phase
space rather than as time-series (Kaasalainen \& Binney 1994, and
references therein).  These surfaces are topologically equivalent to tori,
so we call this the ``torus method''. At present the only three-dimensional
orbits we can construct are axisymmetric ones. In this case each orbit is
characterized by three special isolating integrals, the actions
$J_r,J_l,J_\varphi$. Here $J_r$ is the generalization of the energy $E_R$ of
equation (\ref{OLtwo}); $J_l$ is the generalization of $E_z$ in equation 
(\ref{OL}); and $J_\varphi\,{\equiv}\,L_z$ is simply the angular momentum about
the symmetry axis.

In place of the time-series $({\bf x}_k,{\bf v}_k)$ above we obtain ${\bf x},
{\bf v}$ as Fourier series in the so-called angle variables $\theta_r,\theta_l,
\theta_\varphi$. Along an orbit angle variables have the remarkable property of
increasing linearly in time: $\theta_i(t)\,{=}\,\theta_i(0)\,{+}\,\Omega_i t$,
where the $\Omega_i$ are constants characteristic of the torus. Consequently,
the probability that a star will be observed at any point on its torus is 
uniform in the variables $\theta_i$.

The actions are special in the following important sense. The phase-space
volume occupied by orbits with actions in an elementary cube in integral
space of size $d^3{\bf J}$ is $\int d^3{\bf x}\,d^3{\bf v}\,{=}\,(2\pi)^3
d^3{\bf J}$. In other words, when one uses actions as integrals, equal
volumes in integral (or `action') space correspond to equal volumes in the 
full six-dimensional phase space. Since the distribution function $f({\bf J})$
is a probability density in phase space, $(2\pi)^{-3}f({\bf J})$ is by this 
result also the probability density in action space: $f_i({\bf J})d^3{\bf J}$
is the probability that a star of population $i$ has actions in $d^3{\bf J}$. 
This fact makes it easy to understand the physical implications of any
particular form of the DF $f_i$ of each population. Consequently, when
actions are used as integrals one knows from general astrophysical
considerations in advance of detailed modelling roughly how $f_i$ should
depend on its arguments. In fact, we will specify the functional form of $f_i$
and fit the data by merely adjusting a small number of parameters in the
specified form. 

\section{Dynamical Models of High-$\!$Velocity Populations}
Several high-velocity stellar populations may be distinguished 
spectroscopically, for example, RR-Lyrae stars, blue-horizontal-branch stars,
and low-metallicity subdwarfs.
A dynamical model would associate to each of these sub-components a total
number of stars $N_i$ and a DF $f_i(I_1,I_2,I_3)$ normalized such that $\int\!
d^3{\bf x}\,d^3{\bf v} f_i$ =1. Equipped with a model of this type, one could
replicate within the computer any well-defined observational selection, i.e.\
predict the {\em survey probability distribution\/} for any survey.

For example, Flynn \& Freeman (1993) observed the radial velocities of a sample
of stars that were photometrically selected to be M giants located about
$20{\rm kpc}$ from the Sun towards the south galactic pole. One can evaluate
the model's corresponding survey probability distribution by selecting stars 
from the model according to precisely the same selection criteria and
determining the distribution of radial velocities of these stars, and compare
it with the observed distribution. The model would also predict the distribution
of the stars' proper motions, and these could be compared with any measured 
values.  Notice that in this comparison uncertainties in the distances to 
individual stars, that arose from, for example, a broad distribution in absolute
magnitudes amongst the selected stars, would not play an important role so
long as the absolute-magnitude distribution had been correctly modelled. 

It is easy to see that a model of the type just described unambiguously
predicts the survey probability distributions for a bewilderingly large number
of observational surveys. Indeed, in whatever direction and in whatever 
magnitude range one selects stars of a given photometric or spectroscopic type,
both the radial-velocity and proper-motion distributions are determined once
$\Phi$, $N_i$ and $f_i$ have been chosen.

Of course, not all observations can be used to test the model; some will be
employed to determine $\Phi$, $N_i$ and $f_i$ and there will be nothing to
be learnt by using the model to replicate these particular observations. So
it is important to estimate how many observations will be needed to establish
a model.

The model depends upon a handful of three-dimensional objects: $\Phi({\bf x})$,
and the $f_i(I_1,I_2,I_3)$. If we assume (for the moment) that the Galaxy is 
axisymmetric, then $\Phi$ is specified by the ISM throughout the plane and we
have to guess only how $\Phi$ changes away from the plane. For any such guess
we can determine $f_i$ in a large part of three-dimensional integral space
merely by determining the velocity distribution of each population $i$ near the 
Sun (e.g., May \& Binney 1986). If the velocity distribution can be determined
at one or two other well-chosen {\em points\/} in the Galaxy, we can complete
our knowledge of $f_i$. So it should be possible to predict the outcome of 
infinitely many very different surveys from the results of just a handful of
surveys! Of course, our predictions will only be as good as our guessed form of
$\Phi$. But it is clear that we have very much less freedom in extending this
function away from the plane than we potentially have information from even
a moderate number of surveys.

The reason why even limited observational work leads to more information
than the models have freedom is that Jeans' theorem renders the Galaxy a
three-dimensional object, but one which we are privileged to observe in
six-dimensional phase space. Even when our observations of phase space are
seriously deficient, so that, for example, we determine neither the radial
velocities nor the distances of stars, we still observe the Galaxy in more
than three dimensions. Consequently dynamics renders our observations
highly degenerate, and they become stringent tests of our model.

The simplest, yet rather complicated, dynamical models for the stellar halo
will start by assuming that the Milky Way is both axisymmetric and fully mixed.
The first assumption is certainly incorrect for the centre of the Galaxy, but
this might have only little influence at large galactocentric distances. Also,
as we have learnt at this workshop (Majewski, this volume), the assumption of a
well mixed halo is an over-simplification. However, the size of both effects,
non-axisymmetry and non-mixedness, is unknown and can most easily be measured
by comparing the data to such idealised models.

\section{A Simple Example}
To demonstrate the viability of our Oxford approach and to illuminate the 
construction of our models and the prediction of survey probability 
distributions, we created a simple toy model and computed its predictions
for a NGP proper motion survey.

\subsection{The Gravitational Potential}
In order to simplify the procedure we have chosen a scale free mass distribution
consisting of a spheroid and a disk
\begin{equation} \label{density}
\frac{\rho(R,z)}{M_\odot {\rm pc}^{-3}}
	= 1.4 \times 10^5 (R^2 + [z/0.8]^2 )^{-\gamma/2} 
	+ 5.8 \times 10^5 R^{-\gamma} {\rm sech}^2\frac{z}{0.06R}
\end{equation}
with $\gamma\,{=}\,1.8$ and $R$, $z$ measured in parsec. The spheroid's 
ellipticity is E2, while the disk's vertical structure is given by the usual 
${\rm sech}^2$ model with scale height $z_d\,{=}\,0.03 R$. The parameters are
chosen such that at $R_0\,{=}\,8{\rm kpc}$ we have $v_c\,{=}\,200{\rm km\,
s}^{-1}$, $K_z(1.1{\rm kpc})/(2\pi G)\,{=}\,74 M_\odot{\rm pc}^{-2}$, and
$z_d\,{=}\,240{\rm pc}$ in agreement with observations. 
The local mass density is 0.013 and 0.055 $M_\odot{\rm pc}^{-3}$ for spheroid 
and disk, respectively. The gravitational potential is evaluated by inserting
the ansatz (with polar coordinates $r,\vartheta$)
\begin{equation} \label{potential}
	\Phi(r,\vartheta) = 4 \pi G r^{2-\gamma} \left[ g(\vartheta) +
	2088 \frac{M_\odot}{{\rm pc}^{-3}} \ln\cosh\frac{\cos\!\vartheta}{0.06}
	\right],
\end{equation}
into Poisson's equation and solving for $g(\vartheta)$ using a multipole
expansion. The ansatz (\ref{potential}) is chosen such that already a moderate
number of multipoles gives the potential and forces to high accuracy.

The advantage of choosing a scale invariant mass model is the existence of
scaling relations, which reduce the number of independent dimensions of 
action space to two.

\subsection{A Distribution Function for the Metal-Poor Halo}
It is well known that for the most gravitational potentials surfaces of constant
energy in action space are nearly planes $A\,{\equiv}\,aJ_r\,{+}\,bJ_l\,{+}\,
J_\varphi$ with $a$ and $b$ being functions of $E$. In the scale-free case, 
moreover, the shape of the energy surface is invariant under the scale 
transformation, hence $a$ and $b$ are constant. For our mass model we find 
$a\,{\simeq}\,1.66$, $b\,{\simeq1.16}\,$ and $A\,{\propto}\,E^{(4-\gamma)/
(2[2-\gamma])}$.

An approximate DF depending on energy is obtained with $f(A)$. If $f\,
{\propto}\,A^\eta$, then the system has radial density profile $\rho\,{\propto}
\,\Phi^{3/2+\eta(4-\gamma)/(2[2-\gamma])}$. From this consideration we finally 
choose our stellar-halo DF to be
\begin{equation}
	f_{\rm halo}({\bf J}) = A^{-(6-\gamma)/(4-\gamma)}
	\big[A+L_{\rm circ}(1{\rm kpc})\big]^{2(\gamma-\beta)/(4-\gamma)},
\end{equation}
which yields $\rho\,{\propto}\,r^{-\gamma}$ for $r{\ll}1{\rm kpc}$ and $\rho\,
{\propto}\,r^{-\beta}$ for $r{\gg} 1{\rm kpc}$, where we have taken $\beta\,{=}
\,3.5$.

Replacing $A$ in $f$ by $A'\,{=}\,a'J_r\,{+}\,b'J_l\,{+}\,J_\varphi$ produces a
flattened and/or an\-iso\-tro\-pic model by shifting  stars over energy 
surfaces. For our example we choose $a'\,{=}\,b'\,{=}\,1.2$, which mainly shifts
the stars to slightly higher $J_r$, i.e.\ to radially extended orbits.

\subsection{A Distribution Function for the Stellar Disk}
Unlike the mass model, our model for the distribution of population I stars
will be an exponential disk. Suppose all disk stars were on exactly circular
orbits. Then the disk's DF would be of the simple form
\begin{equation} \label{cold_df}
	f({\bf J}) = f_0(J_\varphi)\, \delta(J_r)\, \delta (J_l),
\end{equation}
and the mass $dM$ with angular momentum between $J_\varphi$ and $J_\varphi\,
{+}\, d J_\varphi$ would be $(2\pi)^3 f_0(J_\varphi)\, dJ_\varphi$. For an
exponential disk, this must equal $M_d/R_d^2 \exp({-}R/R_d) R$ $dR$. Expressing
$R$ by the radius of the circular orbit with angular momentum $J_\varphi$, one
thus finds
\begin{equation}
	f_0 = \frac{M_d}{(2\pi)^3R_d^2}
		\left(\frac{dR}{dJ_\varphi}\right)_{\rm circ}
		R_{\rm circ}(J_\varphi)\;
		\exp\!\left(-\frac{R_{\rm circ}(J_\varphi)}{R_d}\right).
\end{equation}
We can get a vertically and radially warm disk by replacing the product of 
delta functions in (\ref{cold_df}) by (Binney 1987)
\begin{equation}
	\frac{\Omega_r({\bf J}) \Omega_l({\bf J})}{\sigma^2_r \sigma^2_l}
		\exp\left(-\frac{J_r\Omega_r({\bf J})}{\sigma_r^2}
		-\frac{J_l\Omega_l({\bf J})}{\sigma_l^2} \right)
\end{equation}
with the orbital frequencies $\Omega_i({\bf J})$. The functional form of 
$\sigma_l(J_\varphi)$ determines the scale height as function of radius.
Using the observed relation $\sigma_l^2\,{\propto}\,\exp({-}R/R_d)$ and a
similar one for $\sigma_r$, we finally get
\begin{eqnarray}
	f_{\rm disk}({\bf J}) &=& \frac{M_d}{(2\pi)^3R^2_d\sigma_r^2(0)
		\sigma_l^2(0)} \Omega_r({\bf J}) \Omega_l({\bf J})
		\left(\frac{dR}{dJ_\varphi}\right)_{\rm circ}
		R_{\rm circ}(J_\varphi)
		\nonumber \\ &{}& \times
		\exp\left( \frac{R_{\rm circ}(J_\varphi)}{R_d} -
		          \left[ \frac{J_r\Omega_r({\bf J})}{\sigma_r^2(0)}
			        +\frac{J_l\Omega_l({\bf J})}{\sigma_l^2(0)}
		   \right] {\rm e}^{ R_{\rm circ}(J_\varphi)/R_d } \right).
\label{diskDF} \end{eqnarray}
We have chosen the $\sigma_i^2(0)$ to result in $\sigma_u\,{=}\,36
{\rm km\,s}^{-1}$ and $\sigma_w\,{=}\,19{\rm km\,s}^{-1}$ at $R_0$. Note that
this simple form for the DF of a dynamically warm, exponential disk is valid for
{\em any\/} axisymmetric potential.

\subsection{An Example for a Survey Probability Distribution}
A survey probability distribution (hereafter SPD) resulting from our dynamical 
model is easily evaluated by a Monte Carlo technique as follows. 
(i)   Choose $({\bf x},{\bf v})$ randomly from the $f_i$; 
(ii)  assign stellar age, mass, and metallicity according to some simple models 
      for the SFR, IMF, and $\tau$-$Z$ relation;
(iii) use stellar evolutionary models to find absolute magnitude and colours;
      and
(iv)  `observe' the star by computing the apparent observables and applying 
      the selection criteria of the survey to compare with.
Alternatively to using theoretical models of stellar evolution one could use
some colour-magnitude relation as observed, say, for star clusters to assign 
`stars' to the phase space points.

As an example we computed from the above model for halo and disk the SPD of
proper motions in the NGP direction. The SFR was taken to set in 12 and 14 
Gyr ago for disk and halo, respectively, with exponential decay times of 12 
and 2 Gyr. The common IMF was assumed to be of the form $m^{-(x+1)}$ where 
$x\,{=}\,1.35$ above and $x\,{=}\,0.3$ below $0.8M_\odot$ with lower cutoff 
at $0.15M_\odot$. We assumed a gaussian $Z$-distribution whose mean grows 
exponentially in time from $10^{-4}$ initially to 0.04 today. Stellar colours 
and magnitudes were otained using the models of Rezini \& Voli (1981) and 
Maeder (1981 \& 1991). To translate $({\bf x},{\bf v})$ into distance and
proper motion we used $z_0\,{=}\,7$pc and $v_\odot\,{=}\,(9,12,7) 
{\rm km\,s}^{-1}$.

\begin{figure}
\plottwo{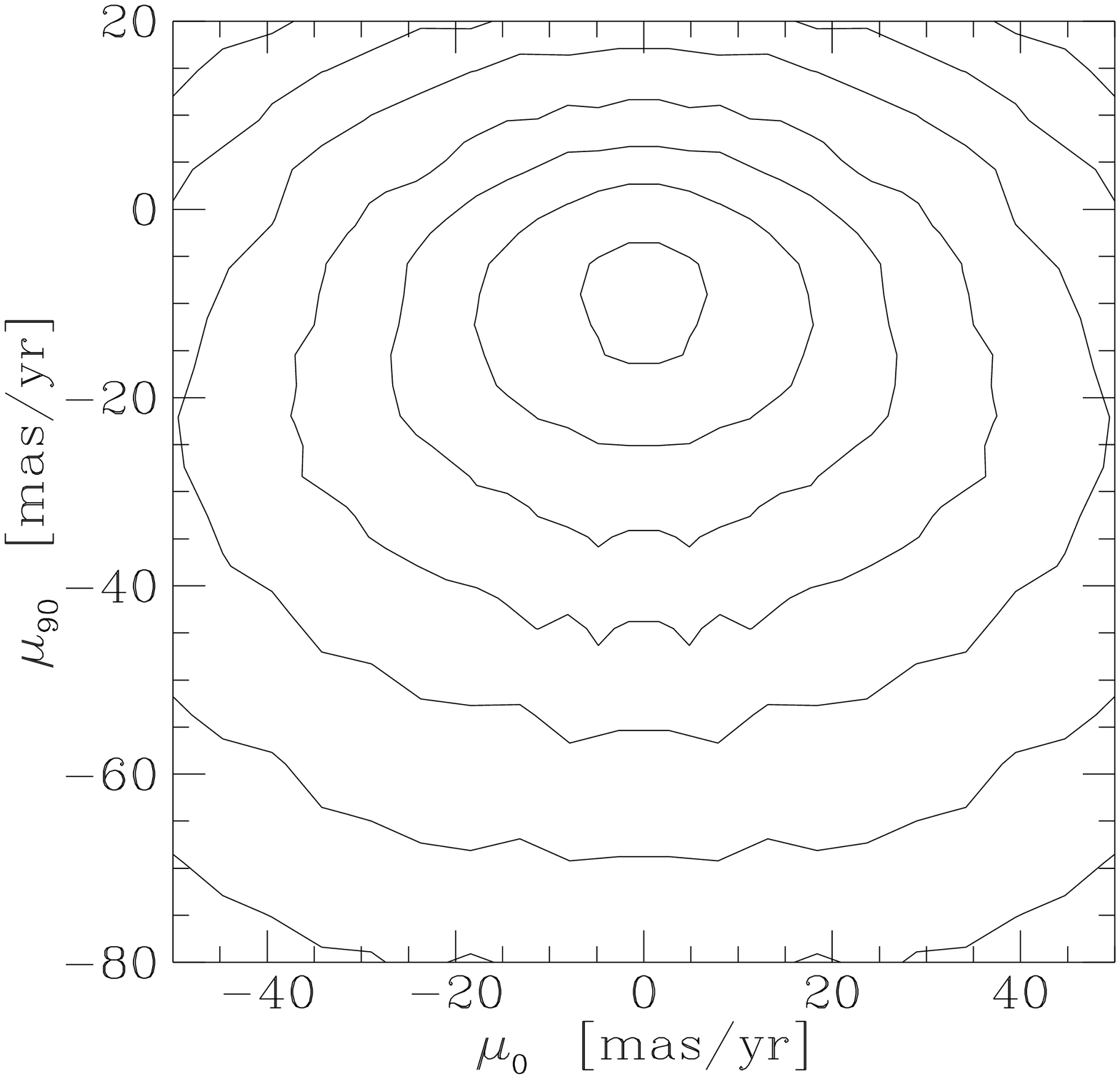}{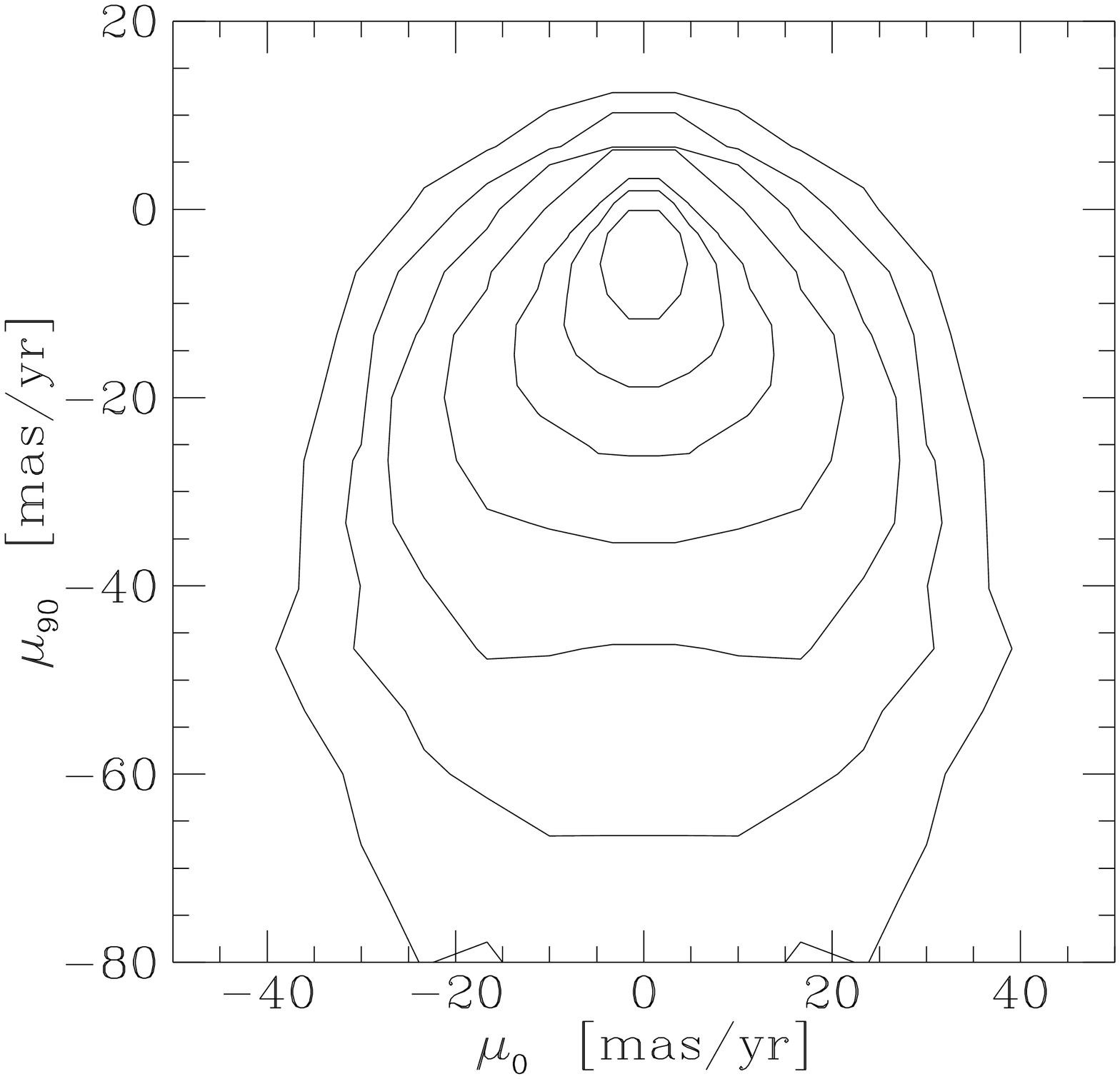}
\caption[]{Contours of the probability for observing proper motions $\mu_0$
	 (in GC direction) and $\mu_{90}$ (in rotation direction) as computed 
	 from our model for a NGP survey with apparent magnitude limit 
	 14$\,{<}\,$B$\,{<}\,$18. The distributions for disk (left) and halo 
	 (right) are shown separately. The contours are spaced by 0.4 dex.}
\label{fig-spds}
\end{figure}
\begin{figure}
\plottwo{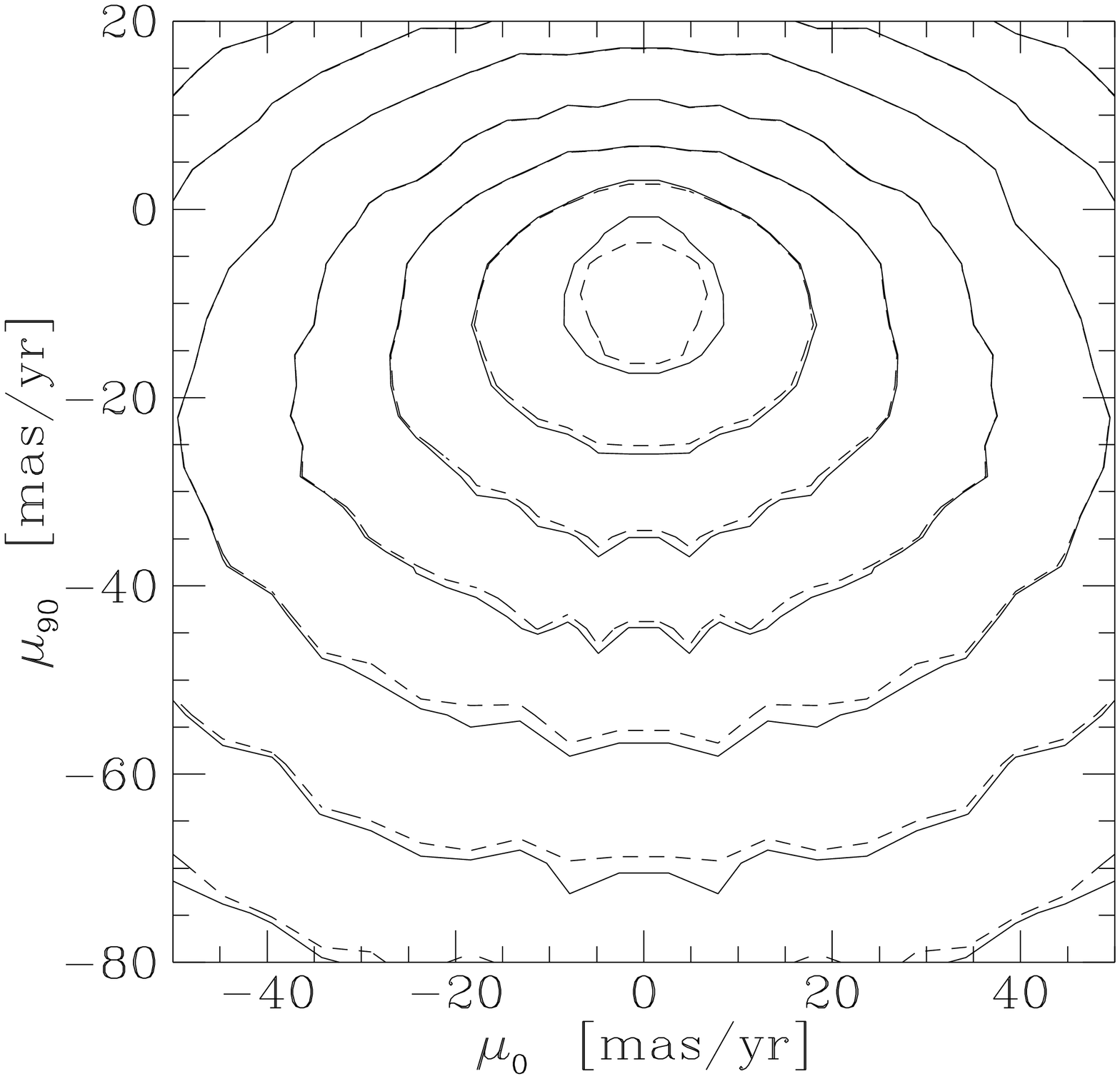}{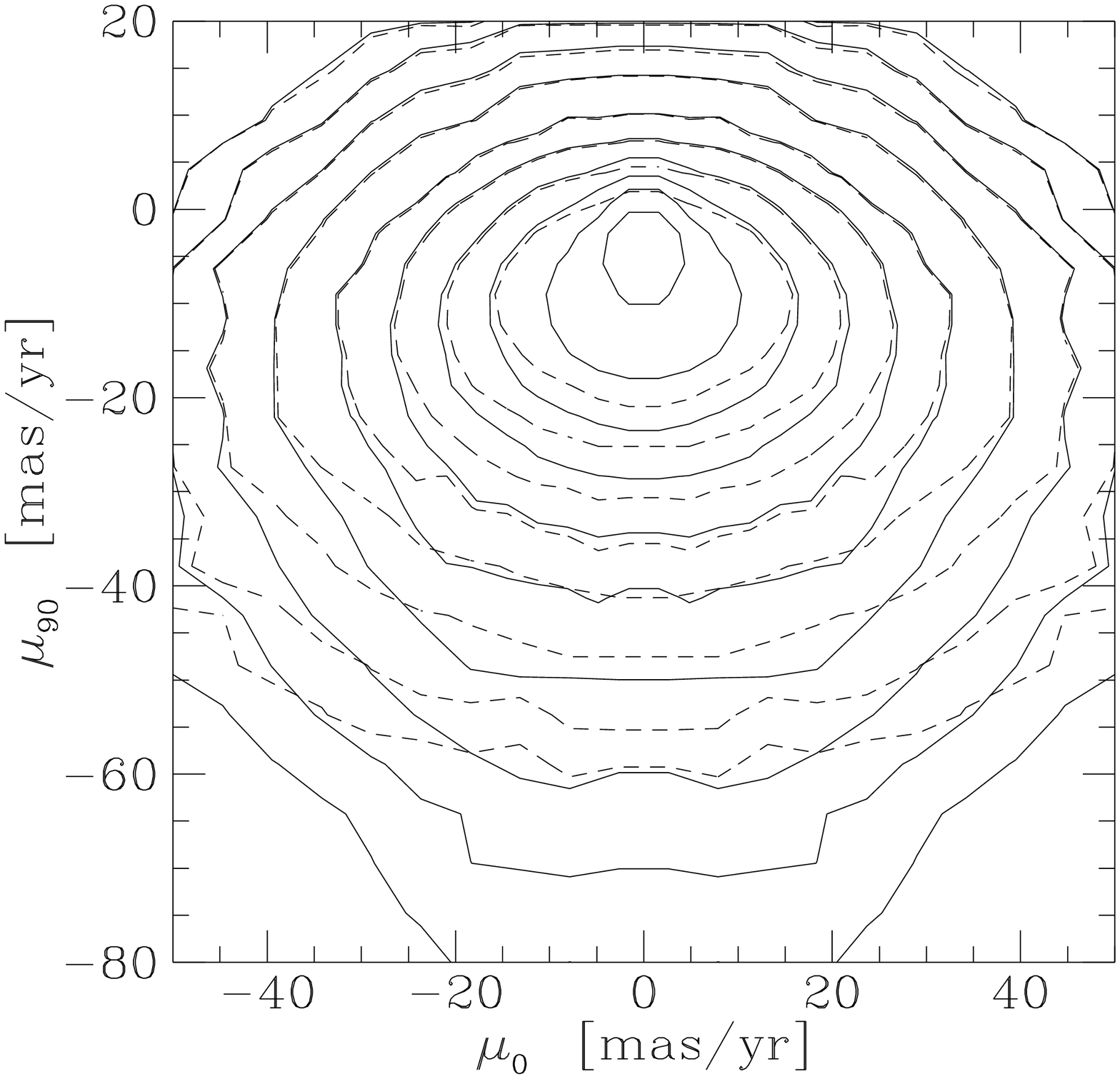}
\caption[]{As Fig.\ {\ref{fig-spds}} but for the joint distribution (solid) of 
	 disk and halo with a local normalization $100\rho_{\rm halo}\,{=}\,
	 \rho_{\rm disk}$. Left and right panel are for
	 14$\,{<}\,$B$\,{<}\,$18 and 18$\,{<}\,$B$\,{<}\,$22, respectively. 
	 The dashed contours show the contribution due to the disk alone.}
\label{fig-jspds}
\end{figure}
The resulting SPD for the disk (Fig.\ \ref{fig-spds}, left) clearly shows
the effects of asymmetric drift: the steady downward shift in the centres
of contours with increasing dispersion (contour size).
This effect becomes much more significant for more local surveys (not shown).
Though non-rotating, the halo SPD (Fig.\ \ref{fig-spds}, right) peaks near
zero proper motion, since most halo stars are too far away to show significant 
proper motion; the few nearby halo stars create the tail at $\mu_{90}\,{<}\,0$.

In the joint distribution (Fig.\ \ref{fig-jspds}, left) the halo stars 
contribute negligibly. Thus, in order to use the halo stars as dynamical 
tracers, one has either to go for deeper surveys (Fig.\ \ref{fig-jspds}, right),
or to disentangle disk and halo stars -- for instance in our model there are
hardly disk stars in the colour-magnitude range B$\,{>}\,$12+10(B${-}\!$V).

\section{Conclusions}
Very considerable advantages flow from modelling the Milky Way in a way that
exploits Jeans' powerful theorem. It is especially useful to model stellar
populations that move far from the plane. What has held up this type of
modelling for decades has been the difficulty of handling orbits that are
constrained by a non-classical integral such as $I_3$. Several viable
approaches to this problem are now available so that it should soon be
possible to compare the new data that will flow from exciting observational
tools such as Hipparcos and the 2dF with fully dynamical models.

We have described some preliminary models based on the Oxford torus technique.
In this method each population is assigned a DF $f_i$ that has a simple
functional form, which describes the population's dynamical characteristics
in astrophysically comprehensible terms. The parameters in $f_i$ are
adjusted so as to optimize the fit between real catalogues and simulated
ones. In principle {\em any\/} catalogue that has well-defined selection
criteria can be simulated.

The modelling process starts by guessing the Milky Way's potential
$\Phi(R,z)$. Significant errors in this guess will lead to discrepancies
between the real and simulated catalogues. Since the fitting process is
strongly over-determined, we are confident that we will be able to determine
$\Phi$ to reasonable accuracy by iteratively adjusting it until the real and
simulated catalogues agree.

\acknowledgments

The authors are grateful to Mikko Kaasalainen for help in creating parts of
the software, and to Uta Fritze-von Alvesleben for making available the stellar
evolutionary tracks in electronically readable form. W.D.\ acknowledges 
financial support by the PPARC.

\end{document}